\documentclass[amssymb,twocolumn,
pra,aps,floatfix,showpacs]{revtex4}

\usepackage{graphicx}  
\usepackage{bm}        
\usepackage{amssymb}   
\usepackage{amsmath}
\usepackage{verbatim}   
\usepackage{blindtext}
\usepackage{subfigure}
\usepackage{upgreek}
\usepackage[usenames,dvipsnames]{color}

\usepackage[pdftex,colorlinks=true,linkcolor=blue,citecolor=blue,urlcolor=blue]{hyperref}

\newcommand{\intensity}[2]{$#1\times10^{#2}\mbox{ Wcm}^{-2}$}

\begin{document}

\title{On the Cooper Minimum in Singly ionized and Neutral Argon}

\author{O. Hassouneh}
\affiliation{Department of Physics, University of Jordan, Amman, 11942, Jordan}
\author{N. B. Tyndall}
\author{J. Wragg}
\author{H. W. van der Hart}%
\author{A. C. Brown}
\affiliation{Centre for Theoretical Atomic, Molecular and Optical Physics, School of Mathematics and Physics, Queen's University Belfast, Belfast  BT7 1NN, United Kingdom}

\begin{abstract}
  We present an analysis of the appearance of the Cooper Minimum in singly
  ionized argon in both the photoionization cross-section (PICS) and
  high-harmonic generation (HHG) spectrum. We employ two 
  computational approaches based on the same R-matrix technique to provide a
  coherent description of the atomic structure of the Ar$^+$ system, finding
  that the PICS and HHG spectrum are affected differently by the inclusion of
  additional residual ion states and the improved description of correlation
  effects. Both the PICS and HHG spectrum possess a clear minimum for all atomic
  structure models used, with the centre of the minimum at 55~eV in the PICS and
  60~eV in the HHG spectrum for the most complete description employed. The HHG minimum is
  systematically shifted to higher energies with respect to the PICS minimum. We
  also find that the initial magnetic alignment (magnetic quantum number) of the
  Ar$^+$ system does not affect substantially the position and shape of the HHG minimum (given a
  sufficiently detailed atomic structure description), but the harmonic yield is
  enhanced by two-orders of magnitude for $M_L=1$ over $M_L=0$. We also perform
  similar calculations for neutral argon, finding that this system is 
  more sensitive to enhancements in the atomic structure description.
\end{abstract}

\pacs{32.80.Rm, 31.15.A-, 42.65.Ky}

\maketitle

\section{Introduction}

Attosecond physics provides a key window on the fundamental 
dynamics of atoms \cite{attosecond_review} and, via the uncertainty
relationships, these dynamics are complimentary to atomic structure. This
complementarity has recently been investigated in the build up of Fano
resonances \cite{fano_buildup}, but is present in other processes. High-harmonic
generation (HHG) has long been established as both the source of ultrashort
light pulses \cite{43as_pulse} and a valuable measurement tool in the guise of
high-harmonic spectroscopy \cite{attosecond_spectroscopy_review}. The {\it dynamic}
HHG process depends on a {\it well-timed} photoabsorption process in a short
laser pulse. This can be
contrasted with a {\it structure-based} photoionization process, which involves light
pulses effectively infinite in extent.

The HHG process has been well described for decades using the simple-man model:
an electron is ionized and accelerated by a strong laser field before
recolliding with its parent ion and emitting its absorbed energy as a
high-energy photon \cite{corkum1993}. The simple-man model provides an intuitive
classical picture for the process, and describes the gross features of the
resulting spectrum of emitted light well, but fails to capture atomic-structure or
multielectron effects \cite{multielectron_atoms}.  

This is particularly troublesome because
probing {\it atomic resonances} with strong-fields has long been a vital function in
attosecond physics \cite{cooper_minimum_hhg_worner, cooper_minimum_hhg_higuet,
cooper_minimum_krypton, multielectron_atoms}. Certain resonances have been shown
to increase the yield of harmonic emission, either in a broad energy range (e.g
via the giant resonance in xenon \cite{multielectron_atoms,
TDCIS_Xe_giant_resonance}) or at the specific frequencies of 
the resonances (e.g. via the window resonances in argon \cite{brown_prl,
window_resonances_hhg}) providing important efficiency gains toward attosecond pulse
generation.

We have shown previously that HHG at the single-atom level is crucially sensitive to the details of
atomic structure of the target system
\cite{ola_multichannel_ne+,brown_ar+} and to the contribution of multiple
electrons \cite{brown_m1,brown_xuvhhg}. One means of
addressing the description of multielectron dynamics is to extend
traditional methods-- based on the single-active-electron approximation-- to
account for electron correlation via correction terms. Recent
attempts at developing such models are strongly suggestive of the crucial role played by multielectron
dynamics in determining HHG emission rates, but are unable to provide a quantitative
description of HHG yields \cite{many_e-_hhg}. 

The need for an accurate, quantitative description of HHG from rare gas atoms
and ions to guide experiment is the motivating factor for several, competing
theoretical approaches.  Several methods have investigated HHG in the
two-electron He atom by solving the fully-dimensional Schr\"{o}dinger equation
\cite{HELIUM, hhg_helium_starace}. However, such methods are not easily extended
to the general multielectron case.  One of the most successful approaches for
general atoms is the Time Dependent configuration-interaction singles (TDCIS)
approach, \cite{TDCIS, TDCIS_spin-orbit,
TDCIS_Xe_giant_resonance,multichannel_hhg}. As the name implies, this method is
restricted to the description of singly excited configurations, and the
calculations are only applicable to closed-shell systems. 
In recent years, several other methods have been proposed which offer various
advantages for the description of strong-field processes in multi-electron
systems \cite{analytical_r_matrix1, analytical_r_matrix2, TDCASSCF,
TDRASSCF-multielectron_atoms, TDRASCI-photoionization,
TDRASSCF-S_many_e_effects}. Among them, R-matrix with time-dependence (RMT) has
emerged as a tractable and flexible approach to general multielectron systems.
Crucially, for the present manuscript, RMT affords selectivity in the detail of
atomic structure included in calculations, allowing for the description of open-shell systems
and multiply excited states. 

RMT is thus the leading method able to describe atomic structure in
sufficient detail to provide direct, \textit{ab initio} solutions of the
time-dependent Schr\"odinger equation (TDSE) for multielectron
atoms or ions. These capabilities have been demonstrated in
the investigation of photoionization of Ne$^{+}$ in combined infrared/XUV pulses
\cite{RMT_rydberg}. The approach has
recently been used to demonstrate the spin-orbit dynamics in the ground state of
carbon with zero and nonzero initial magnetic quantum number \cite{RMT_carbon}. Moreover, the RMT
code has been adapted to include a two-electron finite difference outer region
to allow the modelling of double-ionization
\cite{jack_RMT_two_e,jack_RMT_three_e,tdrm_two_e}. Much recent success in the
application of the RMT method has been to compliment and even preempt cutting
edge experimental techniques such as attosecond transient absorption
spectroscopy \cite{RMT_ATAS}, high-harmonic spectral caustics
\cite{hamilton_two_colour} and extreme-ultraviolet-initiated high harmonic
generation \cite{brown_xuvhhg,clarke_xihhg}.

HHG in ionized noble gas atoms remains a subject of interest for two reasons.
Firstly, the highest harmonics generated in HHG from neutral noble gas atoms
have been attributed to the presence of singly ionized species
\cite{argon+_zepf,argon+_wahlstrom,argon+_gibson}. And secondly, 
the role of multiple, closely-spaced ionization thresholds
represents a useful analogue for molecular systems. 
We have previously addressed HHG in singly ionized neon using a previous
implementation of RMT, demonstrating
the need for the accurate description of multiple ionization thresholds in the
process \cite{ola_multichannel_ne+}. The equivalent process in singly ionized argon is of enhanced interest
because of the Cooper minimum (CM) which we have recently shown to exist
in the harmonic spectrum \cite{clarke_xihhg}. Such minima are the result of a zero in the matrix
element between the $d$ and $s$ continuum waves and the 3$p$-ground state, and have
been widely reported for HHG from Ar atoms following their
long-established appearance in photoionization spectra
\cite{cooper_orig,cooper_minimum_hhg_worner,cooper_minimum_hhg_higuet}. 

The connection between the CM in the photoionization and HHG spectra
has been implicit in many of these previous studies, but no direct comparison
has been made because no method allowed for the simultaneous determination of
the two.
In the present manuscript we present the results of both
photoionization and HHG calculations for Ar$^+$ and investigate therein the appearance
of the CM.

\section{Atomic Structure Descriptions}
\label{sec:structure}
The overarching theoretical framework for our calculations is the $R$-matrix approach
\cite{burke}, wherein configuration space is divided into two regions. In an
inner region, close to the nucleus, we take full account of multielectron
effects, including electron exchange, via a close-coupling with pseudostates
expansion. An electron can be ejected into an outer region by ionization and in
this region the electron is spatially isolated from the residual ion and thus exchange
can be neglected.

Within the $R$-matrix paradigm, then, the
Ar$^+$ states are constructed from a single electron and multi-configuration
Hartree-Fock (MCHF) orbitals for the residual ion states of
Ar$^{2+}$. The calculation includes all final states with a maximum
total angular momentum of 9.

It has been found that there is a strong configuration dependence in
the $3s$ and $3p$ orbitals for the ionic states, as well as for the initial
$3s^{2}3p^{5}$ $^{2}P^{o}$ bound state for Ar$^{+}$. Thus, we include in our
calculations all physical $1s, 2s, 2p, 3s, 3p$ orbitals for the Ar$^{2+}$
description.
Furthermore, to improve the description of the Ar$^{2+}$ ion states, and hence the ionization
process, we also include pseudo-orbitals $4\bar{s}, 4\bar{p}$, $3\bar{d}$ and
$4\bar{d}$ to capture short-range correlation effects not well-described by the
physical orbitals alone.

If only a single set of orthogonal one-electron orbitals are
employed in the calculation, the obtained threshold energy deviates
significantly from the corresponding experimental value. Thus, it is necessary
to include
a large set of doubly-excited configurations to obtain the correct binding
energy.

In this work, we employ a ``3-state" model, including the
three lowest residual ion states-- 
$^{3}P^{e}$, $^{1}D^{e}$ and $^{1}S^{e}$--determined from two different Ar$^{2+}$ models:
\begin{itemize}
\item ``3-CI": including the lowest three CI basis states, whereby,
  $3s^{2}3p^{4}$, $3s3p^5$ and
  $3p^{6}$ are included in obtaining the  wavefunction for each of
  $^{3}P^{e}$, $^{1}D^{e}$ and $^{1}S^{e}$ residual ion states. In this model we add no
  correlation orbitals to the system.  
\item ``9-CI": where we include additional correlation orbitals and allow single and
  double excitations from the physical $3s$, $3p$ orbitals to the $3\bar{d}$
  orbital. Hence, in addition to the above, we include all $3s^23p^{3}3d$,
  $3s3p^{4}3\bar{d}$, $3p^53\bar{d}$,
  $3s^{2}3p^{2}3\bar{d}^{2}$, $3s3p^33\bar{d}^2$ and $3p^{4}3\bar{d}^{2}$ CI basis states in determining 
  the wavefunction. 
\end{itemize}

Including double excitations improves the description of the wave functions near
the nucleus in the residual ion as well as the interaction between the target and
ionized electron at larger distances. This could have an important impact on the
description of the ionization process which impacts on both the PICS and the HHG
spectrum.
The energy spacing between the $^3P^e$ and $^1D^e$ states
is within 20\% of the experimental value in the 3-CI model
and within 5\% in the 9-CI model where we allow for
doubly-excited states by including more correlation orbitals. The same trend is
seen for the $^1S^e$ state (see Tab. \ref{tab:thres}).

\begin{table}[!hbtp]
\caption {Energies of the three ionization thresholds of Ar$^{+}$ with respect to the 
ground state, as calculated in the present scheme and compared to literature
values \cite{argon+_nist}.}
\begin{tabular*}{\columnwidth}{@{\extracolsep{\fill}}cccc}
\hline
\hline
Threshold &      \multicolumn{3}{c}{Energy (eV) }       \\

	        &	Lit.		   & 3-CI & 9-CI\\
\hline
$^ {3}P^{e}$      & 0.00  & 0.00 &0.00\\ 
$^{1}D^{e}$       &1.73&2.04&1.81	   \\ 
$^{1}S^e$          &4.12 & 3.31&4.21 \\                                  
\hline
\end{tabular*}
\label{tab:thres}
\end{table}

In addition, we employ a ``5-state" model, in
which we include the two higher lying residual ion states of $3s3p^{5}$ $^
{3}P^{o}$ and $^ {1}P^{o}$. These higher lying ionization thresholds account for the
emission of $3s$ electrons, whereas, the 3-state models allow the emission of $3p$
electrons only.  
Finally, the ``9-state" model includes an additional two
$^{3}P^{o}$ and two $^{1}P^{o}$ ionic target states.  The $3s^23p^4$~
(${}^1D^e$)~$nd$
states can contain a significant admixture of $3s3p^6$ and when considering
emission of a $3s$ electron, we need to take this mixing into account. Although
the full interplay cannot be accounted for in a tractable calculation, inclusion of
low-lying states may give a first indication of the effects induced by this
interaction. This is afforded by the ``9-state'' model. 


For comparison with Ar$^+$ we also perform a small number of calculations with
neutral argon. The atomic structure descriptions chosen follow a similar
pattern to Ar$^+$. The simplest model, denoted 1S 1-CI (1 state, one
configuration) employs only the $3s^23p^5$ ${}^2P^o$ residual ion state,
effectively restricting ionization to the $3p$ orbital (neglecting the
inner-valence $3s$ orbital). The largest comprises
8 (two ${}^2P^e$ and ${}^2D^e$ and one each
of ${}^2S^e$, ${}^2P^o$, ${}^2F^e$, and ${}^2F^o$) 
residual ion states, and 8 CI basis (8S 8-CI) states which allow all single- and double-excitations
from the physical $3s$ and $3p$ orbitals to the $3\bar{d}$ orbital. We also
employ two models with two residual ion states (${}^2P^e$ and ${}^2S^e$) with 2
CI and 8 CI basis states respectively (2S 2-CI and 2S 8-CI). A complete list of
the residual-ion states and configurations used is shown in Tab.
\ref{tab:models}.

HHG spectra are determined from the time-dependent expectation value of the
dipole operator as extracted from calculations using the RMT code
\cite{RMT,ola_nearIR} while
photoionization cross sections (PICS) for Ar$^+$ are extracted from calculations using the
Dirac Atomic R-matrix Code (DARC) \cite{niall_ar+,DARC_ref,connorb}. 
Calculation of the PICS for neutral argon is performed with the Breit-Pauli (BP)
R-matrix codes \cite{breit_pauli_rmatrix, connorb}. These differ from the DARC
codes in that they apply relativistic
corrections to the Schr\"{o}dinger equation, whereas the DARC codes solve the
Dirac equation. The RMT codes are non-relativstic, but the {\it same atomic
structure description} (atomic orbitals, pseudo-orbitals, configurations) is used as the basis for all calculations performed.
Test calculations with a prototype, relativistic RMT code show that the
effect of the spin-orbit interaction on the HHG spectra for light species such as Ar and Ar$^+$ is minimal,
even though these effects are important for the {\it time-independent}
calculations with the DARC and BP codes.
The use of the R-matrix framework thus allows a direct comparison between
the CM in the HHG spectrum and PICS. However, we note that the HHG spectra are resolved into
contributions from initial states aligned with $M_L=0$ and $M_L=1$, while the
PICS is obtained from a mixed initial state.

\begin{table}[!hbtp]
  \centering
\caption {Atomic structure descriptions for the Ar and Ar$^+$ models used in the
calculations. Model names are derived from the number of residual ion states (S)
and the number of configurations (CI) included in the configuration interaction expansion.}

\bgroup
\def\arraystretch{1.6}
\begin{tabular*}{\columnwidth}{@{\extracolsep{\fill}}cl}
\hline
\hline
Ar$^+$ &  \\
\hline
Model & Residual ion states \\
\hline
3S-$n$CI & ($3s^23p^4$) $^3P^e$, $^1D^e$, $^1S^e$ \\
5S-$n$CI & ($3s^23p^4$) $^3P^e$, $^1D^e$, $^1S^e$, \vspace{-3pt} \\
& ($3s3p^5$) $^3P^o$, $^1P^o$  \\
9S-$n$CI & ($3s^23p^4$) $^3P^e$, $^1D^e$, $^1S^e$, \vspace{-3pt} \\
& ($3s3p^5$) $^3P^o$, $^1P^o$,  \vspace{-3pt}\\
& ($3s^23p^33\bar{d}$) $^3P^o$, $^1P^o$,  \vspace{-3pt}\\
& ($3s^23p^34\bar{s}$) $^3P^o$, $^1P^o$  \\
\hline
Model & Configurations \\
\hline
$n$S-3CI &$3s^{2}3p^{4}$, $3s3p^5$, $3p^{6}$ \\
$n$S-9CI &$3s^{2}3p^{4}$, $3s3p^5$, $3p^{6}$, \vspace{-3pt}\\
& $3s^23p^{3}3\bar{d}$, $3s3p^{4}3\bar{d}$, $3p^53\bar{d}$, \vspace{-3pt}\\
& $3s^{2}3p^{2}3\bar{d}^{2}$, $3s3p^33\bar{d}^2$, $3p^{4}3\bar{d}^{2}$ \\
\hline
&  \\
\hline
\hline
Ar &  \\
\hline
Model & Residual ion states \\
\hline
1S-$n$CI & ($3s^23p^5$) $^2P^o$ \\
2S-$n$CI & ($3s^23p^5$) $^2P^o$ \vspace{-3pt}\\
         & ($3s3p^6$) $^2S^e$ \\
8S-$n$CI & ($3s^23p^5$) $^2P^o$, \vspace{-3pt}\\
& ($3s3p^6$) $^2S^e$, \vspace{-3pt}\\
& ($3s^23p^53\bar{d}$) $^2P^e$, $^2D^e$, $^2F^e$, \vspace{-3pt}\\
& ($3s^23p^54\bar{s}$) $^2P^e$, $^2D^e$, \vspace{-3pt}\\
& ($3s^23p^54\bar{p}$) $^2F^o$\\
\hline
Model & Configurations \\
\hline
$n$S-1CI & $3s^23p^5$ \\
$n$S-2CI &$3s^23p^5$, $3s3p^6$  \\
$n$S-8CI &$3s^23p^5$, $3s3p^6$, \vspace{-3pt}  \\
         &$3s^23p^43d$, $3s3p^53d$, $3p^63d$,\vspace{-3pt}  \\
         &$3s^23p^33d^2$, $3s3p^43d^2$, $3p^53d^2$\vspace{-3pt}  \\
\end{tabular*}
\egroup
\label{tab:models}
\end{table}


\section{Calculation Parameters} For the RMT calculations we employ an 8-cycle,
800~nm laser pulse with a 4-cycle $\sin^{2}$ turn-on and turn-off and peak
intensity of \intensity{4}{14}. The pulse is polarized in the
$z$-direction and is spatially homogeneous. The wavefunction is propagated using
an Arnoldi propagator of order 8 with a time step of 0.01 a.u.. The radial wave
functions of the inner electrons are expanded on a set of 60 11th order
B-splines, using an exponential distribution of knots near the nucleus to
near-linear near the boundary with outer-region. This allows for a numerically
accurate description both close to, and far away from, the nucleus up to the
boundary at $r_{max}=15$ a.u. The grid spacing in the outer region grid is 0.08
a.u. where the radial wavefunction of the outer electron is expanded on a
finite-difference grid out to a radial distance of 2000 $a_{0}$. 

As described in
ref. \cite{brown_helium}, using the non-relativstic Lienard-Wiechert potentials in the far field
limit, the electric field produced by an accelerated charge is 
\begin{equation} 
  \notag
  E(t)=k\left\langle\Psi(t)\left|
  \frac{[p_z,H]}{i\hbar}\right|\Psi(t)\right\rangle + ke  E_{\mathrm{laser}}(t),
\end{equation}
where $\Psi(t)$ is the time-dependent wavefunction of the system, $e$ is the charge of the electron, $z$ is the polarization axis, $k$ is a
constant of proportionality, $p_z$ is the canonical momentum and $E_{\mathrm{laser}}$ is the
electric field of the laser pulse. Expanding the commutator, we obtain
\begin{equation} \notag
  \left \langle \Psi(t) \left| \frac{[p_z,H]}{i\hbar} \right|  \Psi(t) \right \rangle =
  \frac{d}{dt} \left \langle \Psi(t) | p_z | \Psi(t) \right \rangle,
\end{equation}
and it follows that 
\begin{equation}
  E(t) \propto \mathbf{\ddot{d}}(t) =
  \frac{d^2}{dt^2}\langle\Psi(t)|\mathbf{z}|\Psi(t)\rangle,
\end{equation}
where $\mathbf{z}$ is the position operator.
Then, the power spectrum of the emitted radiation is given, up to a
proportionality constant,  by
$|\mathbf{\ddot{d}}(\omega)|^2$- the Fourier transform of $\mathbf{\ddot{d}}(t)$
squared. The dipole acceleration $\mathbf{\ddot{d}}$ cannot however be 
computed easily (except in simple cases such as atomic helium) as this quantity
is prohibitively sensitive to the description of atomic structure at very small
radial distances. Instead, the relationships between acceleration, velocity and
displacement can be exploited to express the harmonic spectrum in terms of the
the dipole velocity and/or length.

Thus in our calculations the harmonic
spectrum is calculated from the time-dependent expectation value of the dipole
operator $\mathbf{D}$:
\begin{equation} \mathbf{d} (t) =  \langle \Psi(t) | \mathbf{D} | \Psi(t)\rangle
  \notag, \end{equation}
  and of the dipole velocity operator $\mathbf{\dot{D}}$:
\begin{equation} \mathbf{d} (t) =  \langle \Psi(t) | \mathbf{\dot{D}} | \Psi(t)\rangle
  \notag, \end{equation}
The harmonic spectrum is then given by 
\begin{equation} S(\omega) \quad = \quad \omega^4 | \mathbf{d}(\omega)|^2 
 \quad= \quad \omega^2 | \mathbf{\dot{d}}(\omega)|^2 
  \notag,
\end{equation}
where $\omega$ is the photon energy and $\mathbf{d}(\omega)$ and
$\mathbf{\dot{d}}(\omega)$ are the Fourier
transforms of $\mathbf{d}(t)$ and $\mathbf{\dot{d}(t)}$ respectively. Consistency between the length
and velocity form spectra is used a test of the accuracy of the RMT
calculations. For the largest atomic structure description used we obtain
agreement within 20\% between the length and velocity form. The spectra shown
here are all of the length form, but for all models
employed, the comparisons and analysis between HHG spectra apply both to the length and
velocity forms.

For the DARC calculations, the R-matrix inner region was set  at 13.28 a.u which
is just sufficient to encompass the diffuse, $4p$ orbital, and 22 continuum orbitals
are used to describe the outgoing electron. An energy mesh of spacing 55 meV was
used to scan the photon energy range between 27eV and 160eV. The PICS is
calculated in Megabarns (1 Mb = $10^{-18} \mbox{ cm}^2$) as
\begin{equation}
  \sigma = \frac{4\pi a_0^2 \alpha\omega}{3g_i} \sum \langle \Psi_f | \mathbf{D}
  | \Psi_i \rangle \notag
\end{equation}
where $a_0$ is the Bohr radius, $\alpha$ is the fine-structure constant, $g_i$
is the statistical weighting of the initial state and $\Psi_i$ and $\Psi_f$ are
the initial and final state wavefunctions respectively. As with the HHG
calculations, the PICS can also be calculated using both the length and velocity
forms of the dipole operator $\mathbf{D}$. Up to 45~eV, agreement is found between the two to
within 20\% and thus all results shown are provided in the length gauge.

\section{Results}

\subsection{Singly ionized Argon}
\label{sec:results}

\begin{figure}[!htp]
\centering
\includegraphics[width=0.45\textwidth]{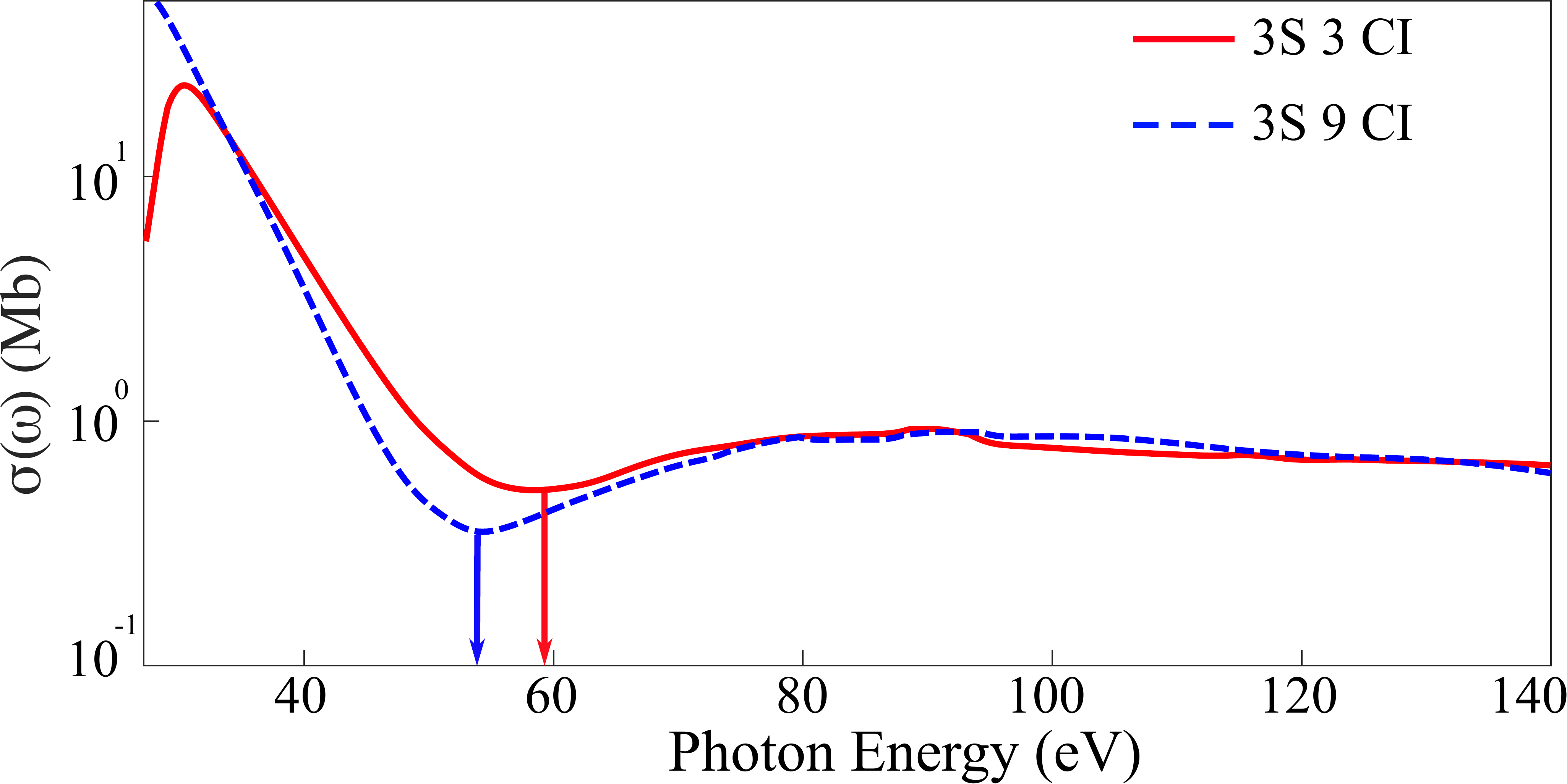}
\caption{The calculated total photoionization cross section,
  $\sigma$, of Ar$^{+}$ with 3-state/3-CI (solid, red
line), and 3-state/9-CI(dashed, blue line) model,  as a function of the
incident photon energy. The position of the Cooper Minimum is denoted by the
vertical lines at 59~eV and 55~eV. The differences between the spectra are
around 10\% in the high-energy region and 50\% around the minimum. The raw data for the figures in this paper can be found
via Ref. \cite{ar+_data}. \label{fig:PICS3sate}
}
\end{figure}  

Figure \ref{fig:PICS3sate} shows the photoionization cross section (PICS) of Ar$^{+}$
for the 3-state model, with the two different CI bases.
For the linearly polarized light pulse here employed, and an initial total magnetic
quantum number $M_L=0$, only $m_l=0$ electrons will be ejected. Thus, this model
allows only for the ejection of a $3p_{0}$ electron, leaving the residual
Ar$^{2+}$ ion in one of the $^{3}P^{e}$, $^{1}D^{e}$ or $^{1}S^{e}$ states.  
The Cooper minimum (CM) is visible in the PICS at 59~eV in the 3-state, 3-CI model.
However, the inclusion of doubly-excited configurations in the 9-CI model leads
to a shift in the position and shape of the CM, giving a much deeper minimum
centred on 55~eV. Measured
from the minimum to subsequent maximum, the CM has a depth of 0.40 Mb in
the 3-CI case, contrasted with 0.59 Mb for 9-CI. The relative change in the PICS
from the additional structure is around 50\% at the CM, and 10\% at high energy
(100~eV).

\begin{figure}[t]
\centering
\includegraphics[width=0.45\textwidth]{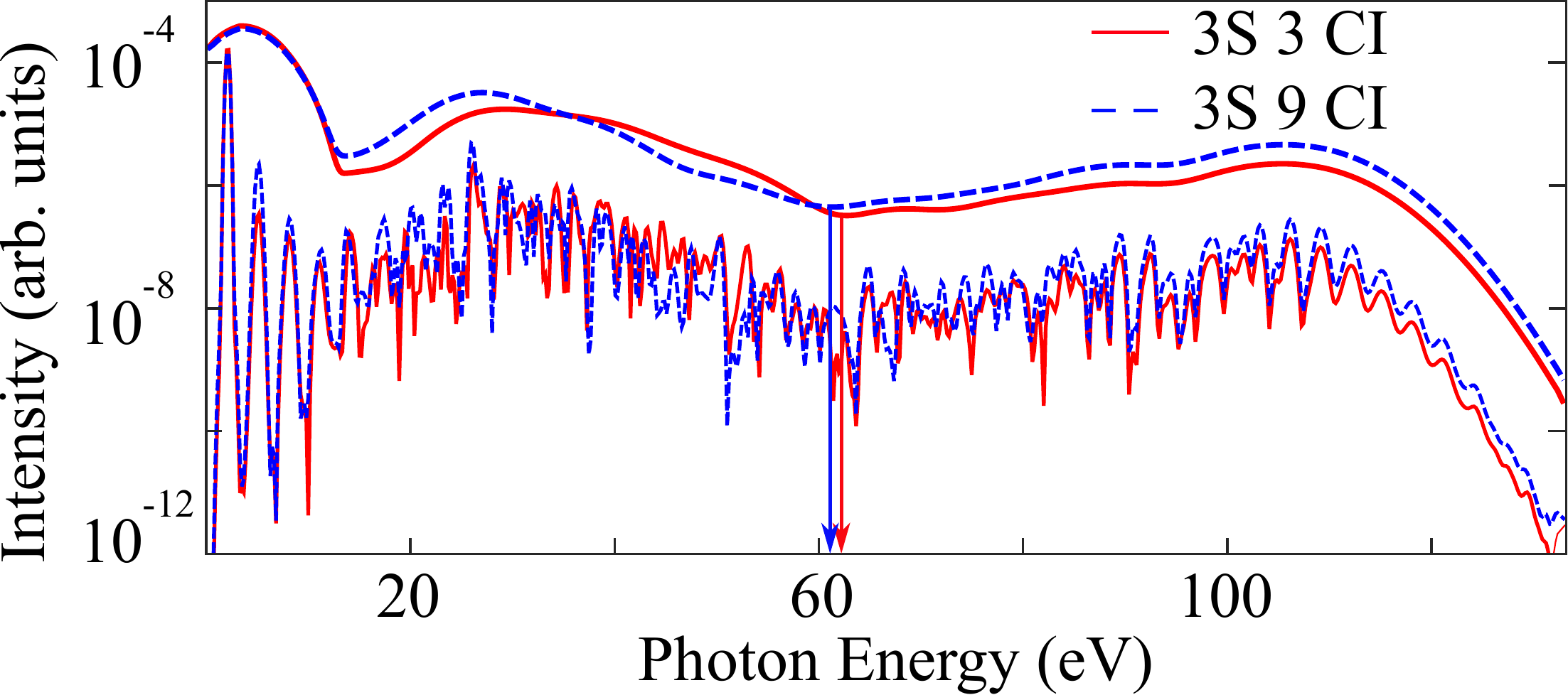}
\caption{ The calculated harmonic spectrum from Ar$^{+}$ at a
laser wavelength of 800~nm and peak intensity of 4$\times 10^{14}$ W
cm$^{-2}$, obtained using the RMT approach for the 3-state/3-CI model 
(solid, red line) and the 3-state/9-CI (dashed, blue line). The upper lines show
the (offset) smoothed spectra which makes clear the slight difference in the
positions of
the Cooper Minimum, which are denoted by the vertical lines at 64~eV (3S 3-CI)
and 62~eV (3S 9-CI). The differences between the spectra are around 40\% at the
minimum and 60\% at the cut-off.
\label{all3targetHHG}}
\end{figure}

Figure
\ref{all3targetHHG} shows the single-atom harmonic spectrum for Ar$^{+}$
generated by a 800~nm pulse, with a peak intensity of \intensity{4}{14}
for the 3-state model, with two different CI bases.
As expected, the CM is 
well replicated in the HHG spectra. 
At energies below the CM the harmonic yield is slightly suppressed in the 9-CI model,
which mirrors the behaviour of the PICS (Fig. \ref{fig:PICS3sate}).
However, at slightly higher energies
the harmonic yield is higher in the 9-CI case, while the opposite is true
for the PICS.
Differences in harmonic yield at energies just above the CM may
cause a shift in the apparent position of the minimum, which appears at 64 eV
for the 3-CI case, but at 62~eV in the 9-CI model. We note that these energies
are higher than the corresponding positions in the PICS (59~eV and 54~eV
respectively),
which is line with previous observations of the CM in neutral argon
\cite{cooper_minimum_hhg_worner,cooper_minimum_hhg_higuet}. However, it is
interesting to note that the PICS and HHG spectrum are affected quite
differently by the enhanced atomic structure: inclusion of the doubly
excited configurations in the 9-CI model has only a small effect on the CM in
the HHG spectrum, but changes its position and depth markedly in the PICS.

Additionally, only small differences (10\%) are observed at high energy in the PICS,
while for the HHG spectra much larger differences (60\%) are obtained around the
cut-off. This can be explained intuitively: in the PICS the high-energy
region is produced by outgoing electrons escaping the ionic potential with
substantial excess energy. In
the HHG spectrum, the highest energies are produced by electrons {\it
recolliding}
with the ion, and the details of atomic structure are thus more pertinent in the
latter case. 

\begin{figure}[b]
  \centering
  \includegraphics[width=0.45\textwidth]{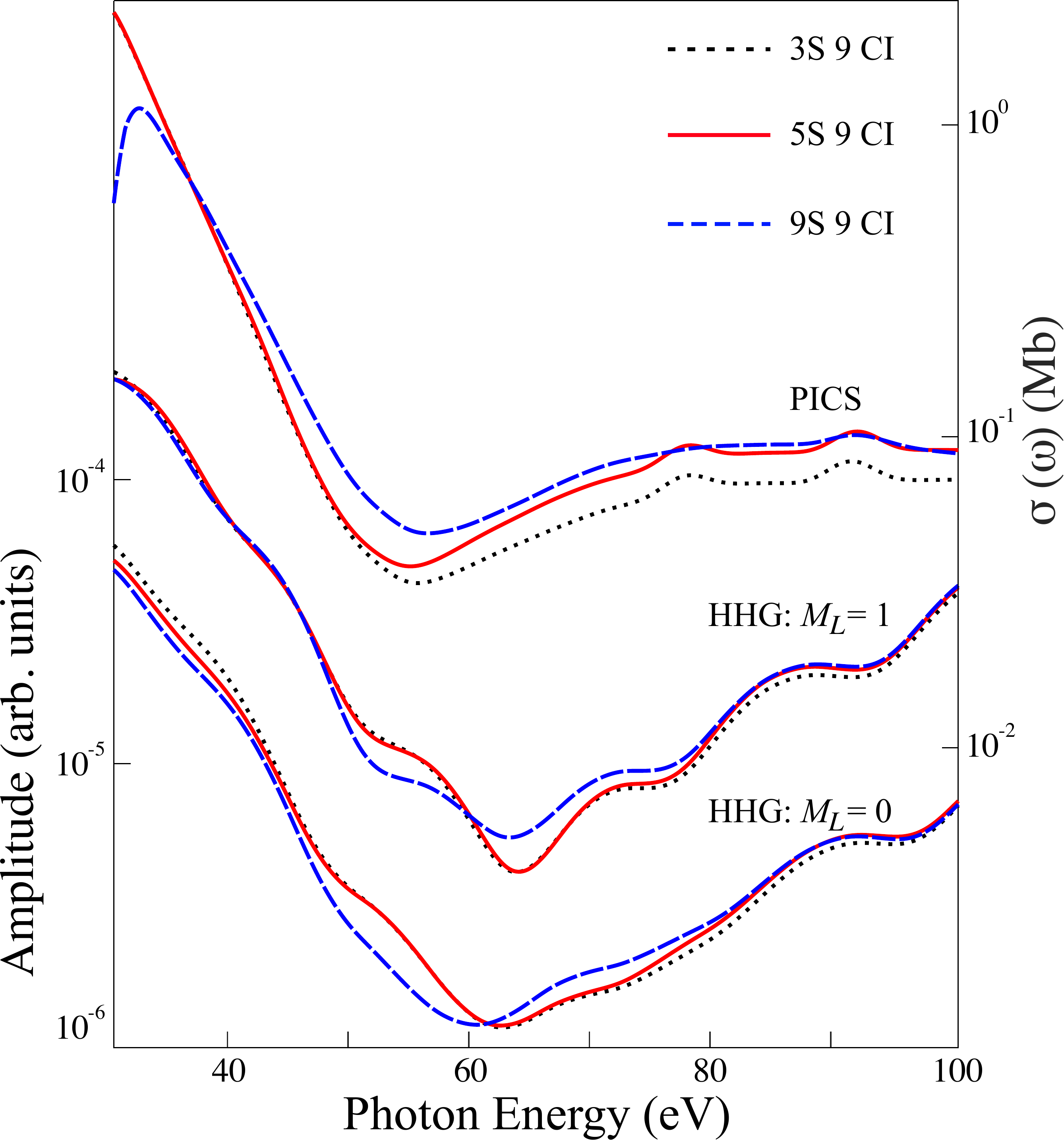}
  \caption{The photoionization cross section (upper) and smoothed HHG spectrum
    in an 800~nm, \intensity{4}{14} field (lower) for Ar$^+$ as calculated
  using the 3-state, 9-CI (black, dotted line), 5-state, 9-CI (red, solid line)
  and 9-state, 9-CI (blue, dashed
  line) models. The HHG spectrum is shown for an initial total magnetic
  alignment of $M_L=1$ (upper) and $M_L=0$ (lower).  \label{fig:residual_comp}}
\end{figure}

Including the $3s3p^5$ residual ion states in the 5-state, 9-CI model allows the
emission of the $3s$ electron. The PICS and HHG spectra obtained from
calculations using both the 3-state and 5-state 9-CI models are shown in Fig.
\ref{fig:residual_comp}.
Although the shape of the PICS is altered somewhat by the inclusion of the $3s$
ionization channels, the position of the CM is changed only by 1~eV, and the HHG
spectrum is not affected by the $3s$ electron at all.

That the CM is robust against the action of the $3s$ electron is to be expected: the CM
is caused by the
relationship between the $3p$ ground state and the $s$ and $d$ continua. The
ionization of Ar$^+$
should, however, be sensitive to all open ionization channels. This
sensitivity is more
pronounced in the PICS, as photoionization of the $3s$ electron by an XUV pulse is
substantially more likely than the tunnel-ionization of the $3s$ in an IR field
as required for HHG. Whereas the photoionization yield of $3s$ at 55~eV is 20\%
of the $3p$ yield, tunnel ionization of $3s$ at 800~nm is less than 1\% of the
$3p$ yield.
Thus while the changes in the PICS might be attributed to the influence of the
$3s$ electron, any direct effect on the HHG spectrum should be negligible. 

Including the further residual ion states in the 9-state models does, however,
illicit more substantial changes in the HHG yield. 
While the direct effect of the these channels on HHG should again be minimal
(tunnel-ionization into the additional channels is 1\% of the total ionization), the
additional structure will have an
indirect effect due to enhanced configuration interaction between $3s$ and $3p$ emission
channels. Changes in the HHG spectrum could be attributed
to these indirect effects.

\begin{figure}[t]
  \includegraphics[width=0.45\textwidth]{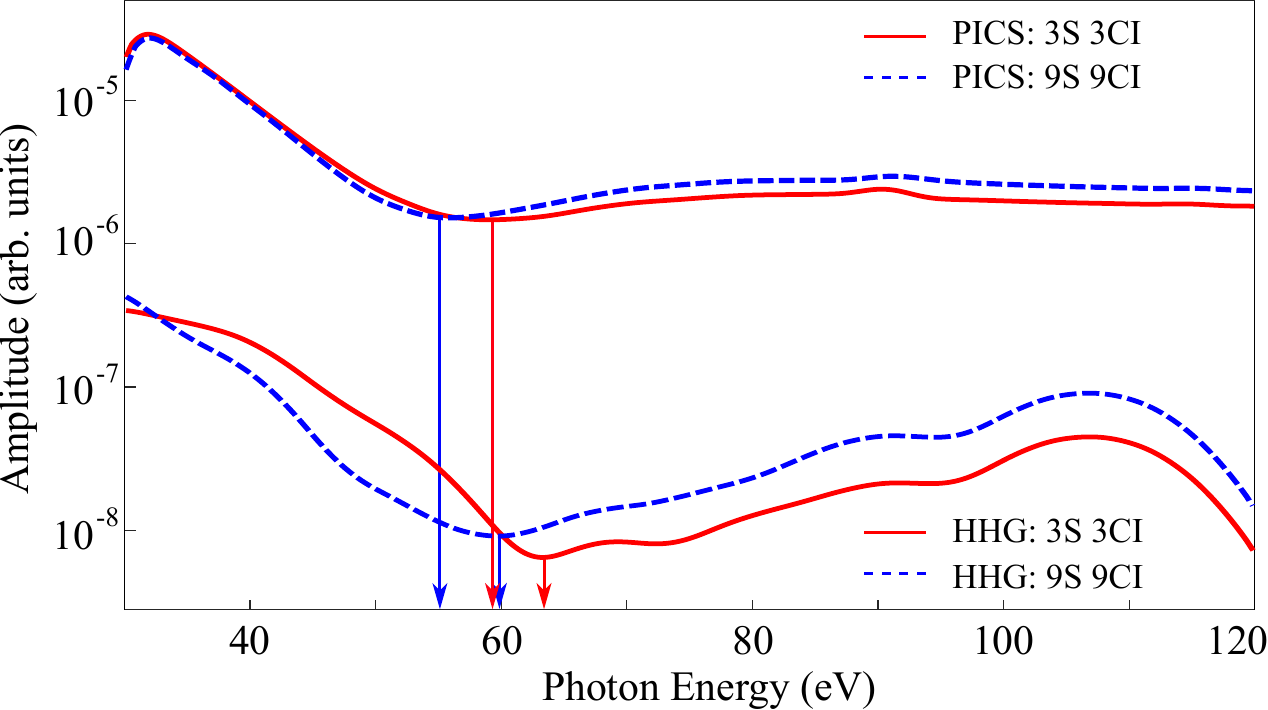}
  \caption{The photoionization cross section (upper) and smoothed HHG spectrum
    in an 800~nm, \intensity{4}{14} field (lower) for Ar$^+$ as calculated
  using both the 3-state, 3-CI (red, solid line) and 9-state, 9-CI (blue, dashed
  line) models. In each case the position of the Cooper Minimum is denoted by
  the vertical line \label{fig:allcomp}}
\end{figure}

Figure
\ref{fig:allcomp} shows the PICS and HHG spectra in both the smallest
(3-state, 3-CI) and largest (9-state, 9-CI) models used. The position of the CM is
shifted down by 4~eV
in both the PICS and HHG spectrum. The enhanced atomic structure
description allows a far greater mixing between the $3s$ and $3p$ emission
channels, which will affect the CM by modifying both the $3p$ ground state of
the Ar$^{2+}$ ion, and the $s$ and $p$ continua of the outgoing or recolliding
electron.

All of this implies that the effect of correlation on the dominant $3p$
ionization channels is substantially more important for HHG than the direct impact of excited
residual ion states. The picture for photoionization is not so clear, where both
the direct impact of additional ionization channels, and the influence of
configuration interaction produce non-negligible changes in the PICS.

We have reported previously that changing the total magnetic
quantum number $M_L$ can have a significant effect on HHG from 
noble-gas ions, including Ar$^+$, albeit for shorter wavelengths \cite{brown_m1,ola_multichannel_ne+}.
Specifically, we have found that for Ar$^+$ initially in the $M_L=0$ state, the HHG
yield was reduced by a factor of four compared to $M_L=1$. For Ne$^+$  the effect
was even more pronounced- with a factor of 26 difference between the two
calculations. Now we assess the effect for an 800~nm laser, allowing us to see the
impact on the CM.

\begin{figure}[tb]
\includegraphics[width=0.45\textwidth]{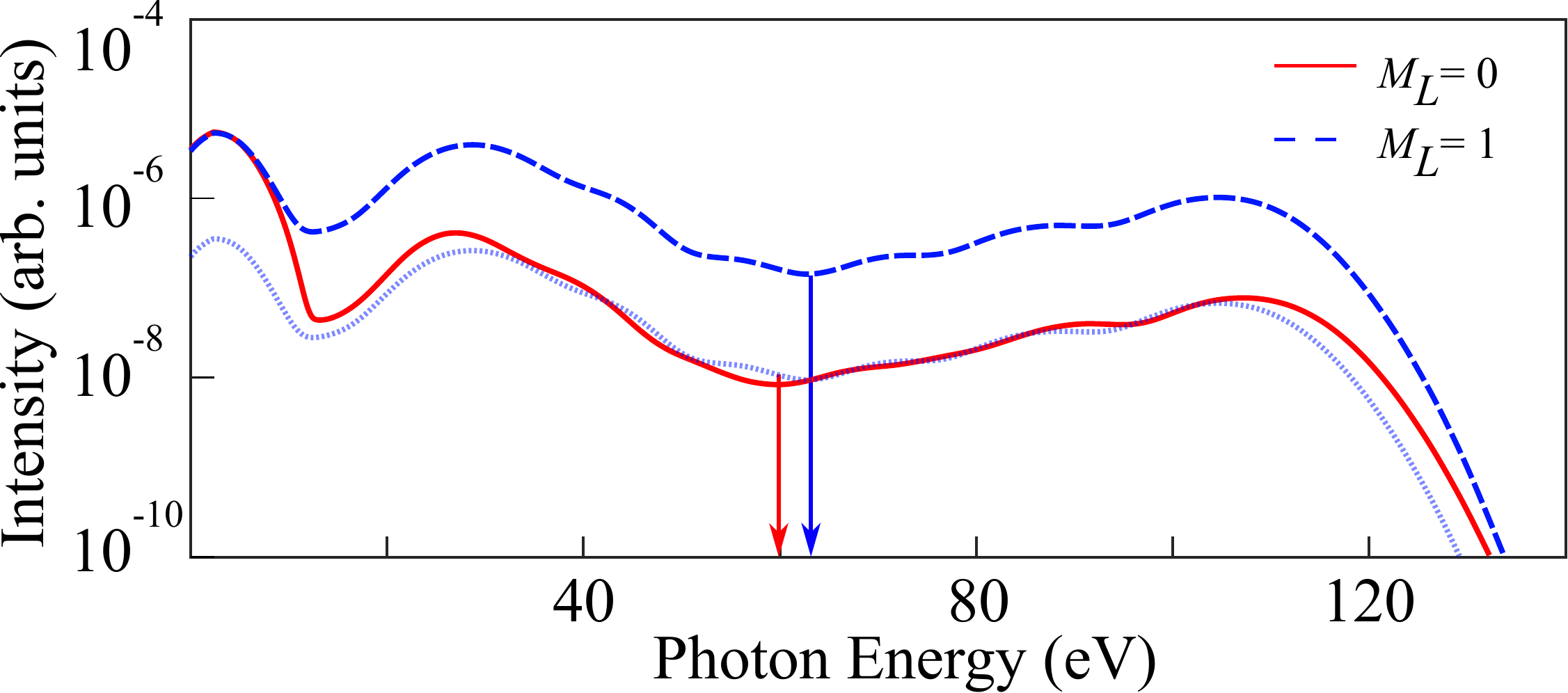}
\caption{ The calculated harmonic spectrum from Ar$^{+}$ at a
laser wavelength of 800~nm and peak intensity of 4$\times 10^{14}$ W
cm$^{-2}$, obtained using the RMT approach for the 3-state/9-CI model 
with an initial total magnetic alignment of $M_L=0$ (solid, red line) and
$M_L=1$ (dashed blue line). The Cooper Minimum, is denoted by the vertical lines
at 64 ($M_L=0$) 
and 63~eV ($M_L=1$). The renormalised yield for $M_L=1$ is shown (light
blue-dotted line) to demonstrate that the shape of the CM is not affected
substantially by the change in magnetic quantum number.
\label{fig:m0m1}}

\end{figure}

Figure \ref{fig:m0m1} shows the HHG spectra for the 9-state, 9-CI model with total
magnetic quantum number $M_L=1$ and $M_L=0$.  The harmonic yield is increased by
two orders of magnitude for $M_L=1$ over $M_L=0$: in line with
Refs. \cite{brown_m1,ola_multichannel_ne+} this reflects an increase in the ionization probability for
$M_L=1$. Despite this substantial increase in yield, the position of the CM is
only slightly altered by the change in the magnetic quantum number, appearing at
60 ($M_L=0$) and 63~eV ($M_L=1$), and in fact, the shape of the minimum is
barely affected at all. The renormalised spectrum for $M_L=1$ almost exactly
overlaps the $M_L=0$ spectrum in the region of the CM. 

\subsection{Neutral Argon}

\begin{figure}[ht]
\includegraphics[width=0.45\textwidth]{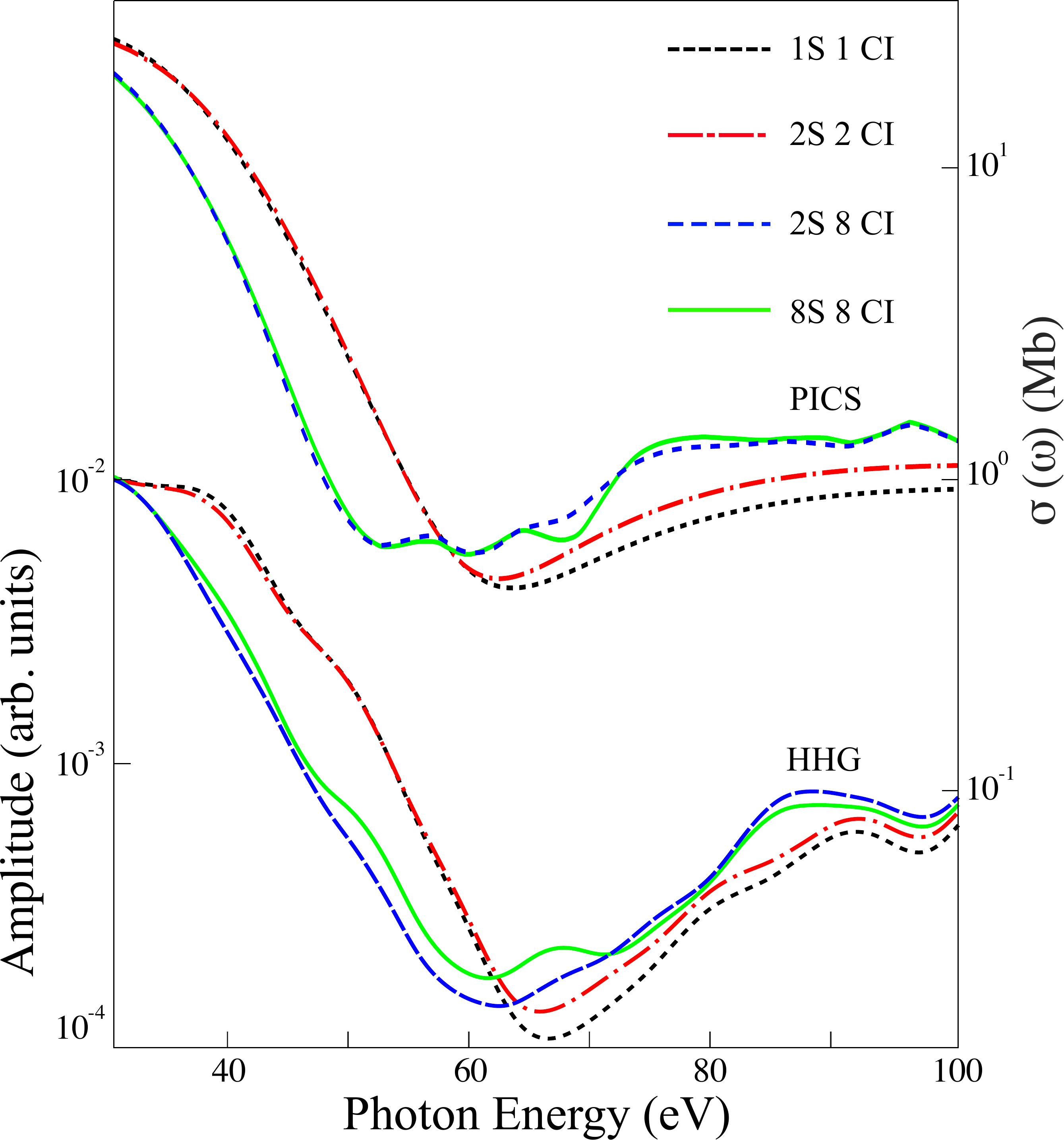}
\caption{ The photoionization cross section (upper) and smoothed HHG spectrum
in an 800~nm,  \intensity{4}{14} field (lower) for Ar 
obtained using the R-matrix approach with the 
1-state 1-CI (black dotted line), 2-state 2-CI (red dot-dashed line), 2-state 2-CI
(blue dashed line) and 8-state 8-CI (green solid line) models.
\label{fig:argon}}

\end{figure}

Figure \ref{fig:argon} shows the HHG spectra and PICS for neutral argon
irradiated by an 8-cycle laser of wavelength of 800~nm and peak intensity
\intensity{4}{14} obtained using the R-matrix approach for the four models described
in section \ref{sec:structure}. The CM is present in these spectra between 50
and 65~eV and-- as was observed for singly-ionized argon-- is shifted
systematically to higher energies in the HHG spectra relative to the PICS. The
inclusion of additional atomic structure causes the position of the minimum to
shift to lower energies for both the HHG spectra and PICS, and adding extra
residual ion states is seemingly less important than the accurate description of
electron correlation which is afforded by the inclusion of more configuration
interaction terms. This is reflected in the fact that the position and shape of
the minimum changes quite sharply between the 2S 2-CI and 2S 8-CI calculations,
but does not change substantially between the 2S 8-CI and 8S 8-CI calculations.

Again, we expect the direct effect of the $3s$ electron to be minimal--
tunnel ionization of the $3p$ electron outweighs that of the $3s$ by a factor of
$10^6$-- and yet the changes wrought by the $3s$ electron's inclusion in both the
HHG spectra and PICS are clearly visible. We can thus attribute this effect to
the indirect effect of correlation between outgoing electron emission channels.
Finally, comparing Fig. \ref{fig:argon} with Fig. \ref{fig:allcomp} it is noticeable that the
inclusion of additional atomic structure has a greater effect on the
appearance of the CM in neutral Ar than in Ar$^+$.

\section{Analysis}

\begin{table}[ht]
\caption{The position 
 of the Cooper Minimum in Ar$^+$ and Ar in both the PICS and HHG spectrum. The
atomic structure descriptors denote the number of residual ion states (S) and
the number of configurations in the configuration interaction (CI) description.
The literature values for the PICS and HHG CM in Ar are shown for comparison and
are taken from ref. \cite{cooper_minimum_pics_samson} and refs.
\cite{cooper_minimum_hhg_worner, cooper_minimum_hhg_higuet} respectively.
\label{tab:CM}}
\begin{tabular*}{\columnwidth}{@{\extracolsep{\fill}}cccc}
  \hline
  \hline
  Atomic & PICS (eV) &\multicolumn{2}{c}{HHG (eV)} \\
  Structure & & $M_L=0$&$M_L=1$ \\
  \hline
  Ar$^+$ & & & \\
  \hline
  3S 3-CI & 59 & 64 & 63  \\
  3S 9-CI & 55 & 62 & 64 \\
  5S 3-CI & 56 & 62 & 63\\
  5S 9-CI & 54 & 62 & 64\\
  9S 9-CI & 55 & 60 & 63\\
  \hline
  Ar & & & \\
  \hline
  1S 1-CI & 63 & 66 & -\\
  2S 2-CI & 62 & 65 & -\\
  2S 8-CI & 53 & 61 & -\\
  8S 8-CI & 52 & 62 & -\\
  Lit.    & 48 & 54 & -\\
  \hline
  \end{tabular*}
\end{table}

Table \ref{tab:CM} shows a summary of the position of the Cooper Minimum for the
various calculations performed in this work. For Ar$^+$ the position of the CM
in the HHG spectrum is systematically higher than in the PICS, and while the gap
ranges from 5~eV (3S 9-CI) to 8~eV (5S 3-CI) the most complete atomic structure
description leads to a gap of 5~eV. For Ar this gap is 10~eV. This is in line with previous studies of
Cooper Minima in other systems-- e.g. experimental measurements in krypton show a shift
of the CM in HHG to higher energies than the PICS\cite{multielectron_atoms}.
However, neutral Ar is a better known case, and some literature values for the
position of the CM in Ar are shown in the table. (We do observe a systematic
shift to higher energies in
our results with respect to the literature values, but note that the relative differences
between the PICS and HHG spectra are consistent with our results)
Higuet {\it et al} found shifts from 50~eV to 54~eV
(theory) 49~eV to 54 eV (experiment) between the PICS and HHG minima
\cite{cooper_minimum_hhg_higuet}. However, these differences could not be
trivially explained by the small difference (2~eV) between the position of the
minimum in the photoionization and photorecombination dipoles.

In the present work however, we can venture some ideas regarding the difference
in the appearance of the CM based on the different
atomic structure descriptions employed in our calculations. Clearly an
improvement in the atomic structure description does not eradicate the
shift and nor should we expect it to, as the shift is reported (for Ar at
least) in various experimental works \cite{cooper_minimum_hhg_higuet,
cooper_minimum_hhg_worner}. Neither do the PICS and HHG spectra respond
uniformly to changes in the atomic structure description-- as noted above
and as is visible from Tab. \ref{tab:CM}. For Ar$^+$ the inclusion of
additional, $3s$ ionization thresholds has a direct 
impact on the shape of the CM in the PICS, but not in HHG, while the enhanced
description of correlation effects, both via the inclusion of additional
configurations and residual ion states, has a similarly marked effect on both
the PICS and HHG spectra. For Ar the situation is similar, although the
inclusion of the $3s$ ionization thresholds has even less of an effect
than for Ar$^+$.

In some sense, this is to be expected-- the HHG spectrum is generated by a
strong field, i.e. is mediated by low energy photons. This will bias the
mechanism towards the ionization of more weakly bound electrons. By contrast,
the PICS is mediated by high energy photons, which can more easily liberate the
inner-valence electrons. Thus, in the Ar or Ar$^+$ picture, the ionization of the $3s$
electrons will be more important in the photoionization process than in the HHG
process. Correspondingly, the inclusion of additional residual ion states, which
allow a better description of the ionization of the $3s$ electrons, will have a more
substantial influence on the shape of the PICS than the HHG spectrum. 

By contrast the inclusion of additional residual ion configurations (i.e. the improved
description of electron correlation) affects the PICS and HHG spectrum
similarly,
shifting the CM position from 59~eV to 55~eV and 64~eV to 60~eV respectively in
Ar$^+$. Furthermore, Fig. \ref{fig:argon} shows that the same qualitative change in the
shape of the minimum is effected by the inclusion of these effects in Ar.

Naive models of HHG in  might neglect all but the ground state target system,
which we have shown previously to have a profound impact on the
resulting HHG yield \cite{brown_prl,brown_ar+}. Here we have demonstrated that even in strong,
long-wavelength laser fields, the details of atomic structure, and correlation
in particular, can influence the appearance of important resonant structures
such as the Cooper Minimum.

\section{Conclusion}
The appearance of the Cooper Minimum in high-harmonic spectra has long been used as a test
case for the theoretical connection between the mechanisms of photoionization
and HHG \cite{multielectron_atoms, cooper_minimum_hhg_higuet}. However, until
recently, the capability to describe atomic processes in strong-fields with an
accurate description of electron correlation was computationally intractable,
and thus a single study addressing both mechanisms in suitable detail has been
elusive. In this paper we have performed a systematic analysis of the effects of
the atomic structure description on the appearance of the CM in singly-ionized
argon in both the PICS and HHG spectrum, finding that each process is sensitive
to slightly different details of the description. Photoionization is more
sensitive to the description of the residual ion states (i.e. the accurate
description of multiple ionization pathways), whereas HHG is
affected more by the details of the electron correlation. 

We find, in concert with studies in other systems, that the CM appears at lower
energies in the PICS than in the HHG spectrum. For the most complete
description used the CM appears at an energy of 55~eV in the PICS and 60~eV in
the HHG spectrum in Ar$^+$ and 52~eV and 62~eV in Ar. We also note that the total magnetic quantum number has
a negligible effect on the position and shape of the CM in Ar$^+$, despite a two-orders-of-magnitude
increase in the harmonic yield for $M_L=1$ over $M_L=0$.

What may seem like insignificant differences in the spectra may in fact become
very important as the attosecond tool-set expands and provides ever more accurate
resolution of atomic structure effects. We have recently reported on XUV
initiated HHG \cite{brown_xuvhhg}, which offers the possibility of stimulating electron correlation
effects with multicolor laser schema (rather than searching for their `natural'
appearance in single-color HHG spectra). In such scenarios the accurate,
time-dependent description of electron-correlation will be absolutely necessary
to analyze experimentally produced spectra and to guide the development of
the theory of attosecond dynamics. To this end, ongoing development of the
R-matrix with time-dependence method will afford capability to investigate 
spin-orbit induced dynamics, the interaction of atomic or ionic systems with
arbitrarily polarised laser pulses, and time-dependent modelling of molecular
systems, all of which are predicated on the ability to describe
atomic (molecular) structure and correlation effects accurately, and resolve
their effects on experimentally observable quantities, as we have done in this paper.

\section{Acknowledgements}
HWvdH and ACB acknowledge financial support from EPSRC, under Grants No.
EP/P013953/1 and EP/P022146/1. OH acknowledges support from The Deanship of Academic
Research and Quality Assurance at the University of Jordan.
This work relied on the ARCHER UK National Supercomputing
Service (www.archer.ac.uk). The data presented in this paper may be found
at Ref. \cite{ar+_data}. The RMT code is part of the UK-AMOR suite, and can be obtained for free from Ref.
\cite{RMT_repo}.

\bibliography{/Users/3048375/Documents/pub/bib_utils/mybib}

\end{document}